\documentclass[sigconf]{acmart}
\usepackage{xcolor}
\usepackage{stfloats}

\AtBeginDocument{%
  }

\begin{document}

\title[A Longitudinal Analysis of Gamification in Untappd]{A Longitudinal Analysis of Gamification in Untappd:\\ Ethical Reflections on a Social Drinking Application}


\author{Jefferson Seide Molléri}
\affiliation{%
  \institution{Kristiania University College}
  \city{Oslo}
  \country{Norway}}
\email{jefferson.molleri@kristiania.no}

\author{Sami Hyrynsalmi}
\affiliation{%
  \institution{LUT University}
  \city{Lappeenranta}
  \country{Finland}}
\email{sami.hyrynsalmi@lut.fi}

\author{Antti Hakkala}
\affiliation{%
  \institution{University of Turku}
  \city{Turku}
  \country{Finland}}
\email{ajahak@utu.fi}

\author{Kai K. Kimppa}
\affiliation{%
  \institution{University of Turku}
  \city{Turku}
  \country{Finland}}
\email{kai.kimppa@utu.fi}

\author{Jouni Smed}
\affiliation{%
  \institution{University of Turku}
  \city{Turku}
  \country{Finland}}
\email{jouni.smed@utu.fi}

\renewcommand{\shortauthors}{Molléri et al.}

\begin{abstract}
  This paper presents a longitudinal ethical analysis of Untappd, a social drinking application that gamifies beer consumption through badges, streaks, and social sharing. Building on an exploratory study conducted in 2020, we revisit the platform in 2025 to examine how its gamification features and ethical framings have evolved. Drawing on traditional ethical theory and practical frameworks for Software Engineering, we analyze five categories of badges and their implications for user autonomy and well-being. Our findings show that, despite small adjustments and superficial disclaimers, many of the original ethical issues remain. We argue for continuous ethical reflection built embedded into software lifecycles to prevent the normalization of risky behaviors through design.
\end{abstract}




\maketitle

\section{Introduction}

Untappd is a social networking application that gamifies beer drinking by allowing users to check in beverages, earn badges, and share their activity with friends~\cite{mather2010untappd}. As a platform that blends social interaction, behavioral incentives, and data tracking, it exemplifies how gamified software can shape everyday practices and legitimize certain behaviors through design. 

In such systems, design choices are never value-neutral. They include assumptions about what constitutes achievement, engagement, and enjoyment~\cite{kim2016more,Hyrynsalmi2017TheDS}. When the underlying activity involves potential health or social risks (as in alcohol consumption) gamification raises ethical concerns~\cite{arora2021ethics}. The Untappd case offers an opportunity to explore how ethics are designed in gamified applications, and their implications for software design and evolution.

Our study consists of two complementary phases conducted five years apart. In 2020, we identified ethically grey areas of gamification, such as badges that encourage high-frequency or high-quantity alcohol consumption. The current phase, conducted in 2025, expands our exploratory findings with a longitudinal analysis informed by netnographic principles. We examine how design changes and community discourses have shifted since the original analysis.

We address the following research questions:

\begin{enumerate}
    \item[\textbf{RQ1}] \textit{How have the design and implementation of gamification in Untappd evolved between 2020 and 2025?}
    We are particularly interested in software design choices that have led to changes in gamification features such as badges, streaks, and nudges.
    
    \item[\textbf{RQ2}] \textit{What ethical challenges emerge from Untappd's gamification when examined through a multi-perspective lens?}
    We focus on individual well-being, community norms, regulatory constraints, and software design ethics.
\end{enumerate}

By addressing these questions, we contribute to Software Engineering (SE) research on responsible, experience-driven design, and extend the literature on gamification ethics. Our intention is not to label Untappd as inherently unethical, but to illustrate a dark side of software design, and argue on a need to embed ethical safeguards when gamification is deployed in socially sensitive domains.

\section{Background and Related Work}

\subsection{Gamification and Ethics}

Gamification is the process of bringing game design elements, such as points, levels, or badges to non-game contexts to increase motivation and/or sustain user engagement~\cite{deterding2011game}. Usually, gamification is used to motivate otherwise arduous and possibly boring tasks, but becomes ethically questionable when it nudges users towards hazardous or socially irresponsible behaviours.

The extant literature discusses ethical problems and considerations of gamification itself~\cite{hyrynsalmi2018road} and its possible solutions~\cite{hyrynsalmi2017gamification, kim2016more, sicart2015playing}. The ethical discourse is often motivated by Star Wars inspired division to the \textit{`Bright Side'} and \textit{`Dark Side'}~\cite{andrade2016bright, callan2014avoid, hyrynsalmi2017dark}. Yet, the categorization into a virtuously good and another unethically evil side, is both naïve and unproductive. Later research has stressed the importance of examining the `grey areas', i.e. contexts where well-intentioned design may still produce harmful outcomes~\cite{hyrynsalmi2017shades,shahri2019engineer}.

Common ethical risks include manipulation, exploitation, and habit formation~\cite{kim2015gamification,Hyrynsalmi2017TheDS}. Frameworks have been proposed to guide ethical gamification design ~\cite{marczewski_ethics_2017,hyrynsalmi2017gamification}, but their applicability remains uneven, especially in complex social platforms. A systematic review~\cite{hamari2014does} highlighted negative consequences such as addictive use, over-competition, or compromised privacy.

Designing gamification features in such ecosystems is complex and resource-intensive~\cite{piras2017gamification}. Studies emphasize the need to integrate ethics into design practices, for instance through frameworks such as Inclusive Gamification, which supports software engineers in identifying individual risks during development~\cite{elsalmy2023inclusive}.

While existing research demonstrates strong theoretical frameworks (e.g.~\cite{elsalmy2023inclusive,alidoosti2021ethics,gordon2020incorporating}), practitioners still face significant challenges in designing ethically aware gamified systems~\cite{karim2017ethical}. It requires that software engineers identify potential risks earlier, document value trade-offs, and embed ethical validation throughout the lifecycle.

\subsection{Untappd Social Drinking App}

Untappd is a mobile social network that enable users to check in and rate beers, using geolocation to tag venues. Each check-in generates data that are shared publicly or among friends~\cite{mather2010untappd}. 

Research have shown how Untappd creates commercial and cultural value, although evidence is primarily observational rather than experimental~\cite{chorley2016pub,santala2017making,ulander2022value}. Chorley et al.~\cite{chorley2016pub} analyzed Untappd’s check-in data to explore social drinking behaviors, illustrating its potential as a proxy for alcohol-related cultural practices. Santala et al.~\cite{santala2017making} investigated how Untappd data supported urban planning decisions. More recently, Ulander~\cite{ulander2022value} examined Untappd from a value co-creation and co-destruction perspective, noting how the platform enables breweries to engage customers while reinforcing problematic patterns of consumption. 

From a user perspective, Untappd's main features are:

\begin{itemize}
    \item Beer check-ins: log and rate beers with photo, timestamp, and location.
    \item Badges and achievements: gamified rewards for frequency, beer-style, location, or event-based drinking.
    \item Social networking: connect with friends, toast (like) and comment on check-ins.
    \item Beer ratings and reviews: scoring and written reviews visible to the community.
    \item Venue integration: brewery and bar profiles, menus, and event promotions.
    \item Statistics and history: personal beer log, trends, and consumption summaries.
\end{itemize} 


Among these, badges\footnote{A list of Untappd badges is available at https://untappdbadges.home.blog/list/.} are the platform’s most distinctive feature, serving as digital achievements awarded to users for specific drinking activities. The platform allows users to view a list of badges they have earned and compare it with friends.


Most app updates and user discourses in social media revolve around new badges and achievements. While Untappd's community guidelines and policies address moderation and privacy, they provide little guidance regarding responsible use or persuasive design. The contrast with less gamified competitors such as BeerAdvocate highlights Untappd's unique combination of playfulness and potential risk, making it a compelling case for ethical analysis.

Untappd does not provide platform-wide aggregated user data, such as on how many users have achieved a particular badge. The Untappd For Business API\footnote{https://docs.business.untappd.com/} allows companies to retrieve analytics for specific locations only. Some third-party tools~\cite{hoffman_webbreacheruntappdscraper_2025,sjaakbanaan_datavisualised_2025} can retrieve badge information, but they are limited to what Untappd exposes publicly.

\section{Research Method}

We adopted a longitudinal interpretive case-study approach~\cite{mcleod2011qualitative,walsham1995interpretive} drawing on netnographic principles~\cite{kozinets2019netnography} about cultural meaning and digital trace interpretation. 

\subsection{Exploratory Phase (2020)}

The first phase explored the ethical dimensions of gamification through philosophical argumentation and interpretive analysis~\cite{koskinen2017ecosystem, stahl2014computer, walsham1995interpretive}. Four researchers independently engaged with the application over a two-month period. Usage patterns varied: some researchers checked in beers as part of their daily drinking routines, while others simulated usage by recording multiple beers in short bursts or at unusual venues.

Data sources included researcher-generated usage logs, screenshots of badges, and collective notes on ethically questionable features. Observations were categorized, discussed, and consolidated into five thematic groups of problematic badges (see Section \ref{sec:findings}). Each category was then examined through virtue-ethical, utilitarian, and deontological lenses to identify misalignments between moral reasoning and gamified behaviour. This phase served to map potential ethical gray areas in the platform’s design.

In the second phase we revisited Untappd to assess how its design and community dynamics had evolved. This research expanded the dataset and analytic scope by integrating digital cultural analysis elements inspired by netnography~\cite{kozinets2019netnography}. Specifically, we treated the platform as a cultural space where user practices and design choices intersect. Our data collection combined:

\begin{itemize}
    \item Systematic walkthroughs documenting current badge structures and design changes.
    \item Official documentation and developer blog posts.
    \item App‐store records, including version histories, update notes, and user reviews.
    \item Public discussions on social platforms such as Reddit.
    \item Comparison with related apps (e.g., BeerAdvocate, HomebrewTalk) and open GitHub projects using Untappd data.
\end{itemize}

\subsection{Analytical Approach}
\label{sec:ethical-framework}\begin{table*}[!hb]
\centering
\caption{Ethical roadmap of Untappd badges using the EASE framework \cite{aydemir2018roadmap}.}
\begin{tabular}{p{1.8cm} p{3.2cm} p{2.8cm} p{2.8cm} p{5.5cm}}
\toprule
\textbf{Badge Type} & \textbf{Subject} & \textbf{Ethical Value(s)} & \textbf{Threatened Object} & \textbf{Lifecycle Implication} \\
\midrule
\textbf{ABV} & Gamification rewarding high-ABV check-ins & Autonomy, safety, well-being  & Individual user well-being, public interest & \textit{Articulation} and \textit{Validation} should include stakeholder value of `responsible drinking.' \\ [4pt]
\textbf{Frequency} & Repetitive streak and progress features & Autonomy, responsibility, sustainability & Individual user autonomy and well-being & \textit{Verification} should ensure no exploitation of compulsive behavior. \\ [4pt]
\textbf{Quantity} & Cumulative tracking of check-ins & Responsibility, transparency, well-being & Individual and social well-being & \textit{Implementation} should include transparent feedback loops on consumption; \textit{Validation} should review societal impact. \\ [4pt]
\textbf{Time of Day} & Context-sensitive badges for unusual hours & Autonomy, responsibility & Social norms, user autonomy & \textit{Verification} should assess whether contextual triggers (e.g., morning use) violate responsible-use. \\ [4pt]
\textbf{Venues} & Location-based check-ins & Privacy, safety, transparency & User location data & \textit{Articulation} and \textit{Specification} must ensure explicit consent and privacy protection. \\
\bottomrule
\end{tabular}
\label{tab:ease_badges}
\end{table*}

\begin{table*}[!hb]
\centering
\caption{Conceptual evaluation using Caragay et al.’s concept-based framework \cite{caragay2024beyond}.}
\begin{tabular}{p{1.8cm} p{4.6cm} p{4.6cm} p{5.5cm}}
\toprule
\textbf{Badge Type} & \textbf{Expected Concept} & \textbf{Observed Concept} & \textbf{Deviation and Ethical Consequences} \\
\midrule
\textbf{ABV} & Reward exploration in beer tasting & Rewards stronger, high-ABV drinks & Misalignment between user expectation and actual incentive, leading to health risks. \\[4pt]
\textbf{Frequency} & Reward consistent engagement & Encourages daily or streak-based check-ins & Normalizes compulsion; manipulates user autonomy to maintain streaks. \\[4pt]
\textbf{Quantity} & Reward milestones or new experiences & Rewards large-quantity consumption & Prioritizes engagement metrics over user well-being; turns consumption into a visible status marker.\\[4pt]
\textbf{Time of Day} & Celebrate leisure and social events & Rewards unusual hours drinking & Breaks normative expectations; reframes inappropriate contexts as acceptable. \\[4pt]
\textbf{Venues} & Promote exploration and businesses & Encourages check-ins in sensitive or unsafe places & Violates privacy and safety expectations. Blurs boundaries between engagement, advertising, and surveillance.\\
\bottomrule
\end{tabular}
\label{tab:concept_badges}
\end{table*}

Our ethical analysis combined major ethical theories with SE perspectives. Specifically, we used:

\begin{itemize}
    \item Virtue ethics~\cite{aristotle2020nicomachean}, utilitarianism~\cite{mill1962utilitarianism}, and deontology~\cite{kant1970kant} to interpret moral implications of gamified behaviors;
    \item The Ethically-Aware Software Engineering (EASE) approach~\cite{aydemir2018roadmap} to locate ethical risks such as autonomy, privacy, and sustainability across the software lifecycle; and 
    \item Caragay et al.'s concept-based framework~\cite{caragay2024beyond} of software design ethics to identify divergences between intended expectation and perceived outcome.
\end{itemize}

We operationalized frameworks by mapping Untappd’s five badge categories against their conceptual dimensions. Tables \ref{tab:ease_badges} and \ref{tab:concept_badges} summarize these mappings according to the EASE framework \cite{aydemir2018roadmap} and Caragay et al.'s \cite{caragay2024beyond} concept-based framework, respectively.

To guide interpretation, we employed a four-dimensional perspective linking badge mechanics with ethical concerns from micro- to macro-levels, as follows:

\begin{itemize}
    \item \textit{Individual well-being.} How badges and streaks affect users' health or autonomy.
    
    \item \textit{Community norms.} How collective play reinforces social expectations around drinking.
    
    \item \textit{Commercial and regulatory context.} How gamification intersects with marketing and legal boundaries.
    
    \item \textit{Philosophical and design ethics.} How design metaphors and reward systems encode moral assumptions.
\end{itemize}

\section{Findings}
\label{sec:findings}

Our findings are structured around the five categories of badges that best illustrate Untappd's gamification design: (1) alcohol-by-volume, (2) frequency of consumption, (3) quantity, (4) period or time of day, and (5) venues. Comparing data from 2020 and 2025 allowed us to trace how each category evolved and what ethical gray areas persisted in the platform's design.

\subsection{Untappd's Current Status}

By 2025, Untappd remains popular, with between 8 and 9 million users and nearly a billion recorded check-ins. The app continued to evolve toward a more modern mobile UI, business-friendly toolset, and broader beverage discovery platform. Key changes from 2020 to 2025 include a Dark Mode, navigation redesign, richer business tools, and enhanced data for users and venues alike.

Although the interface has been modernized, long-standing issues reported by the community persist. Social interaction is limited to basic check-ins and toasts. In 2021, Untappd blocked access to their public API but Untappd for Business API is still maintained and documented. 

Badges are still the core feature. In our current observations, many event-based badges have been removed, yet new ones appear regularly. Newer badges were introduced, such as such as The Century Club (logging the same beer 100 times) or high-volume achievement levels over 15,000. Some badges now include short humorous disclaimers, but the overall reward structure still celebrates repetitive and high-intensity drinking.

\begin{figure*}[!hb]
  \centering
  \begin{minipage}[t]{0.3\textwidth}
    \vspace{0pt}
    \includegraphics[width=\linewidth]{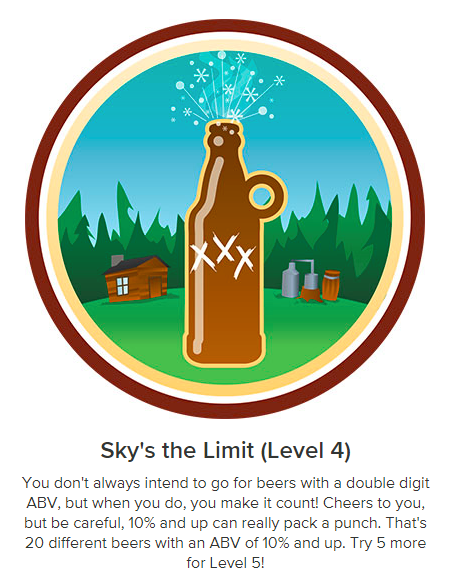}
  \end{minipage}\hfill
  \begin{minipage}[t]{0.3\textwidth}
    \vspace{0pt}
    \includegraphics[width=\linewidth]{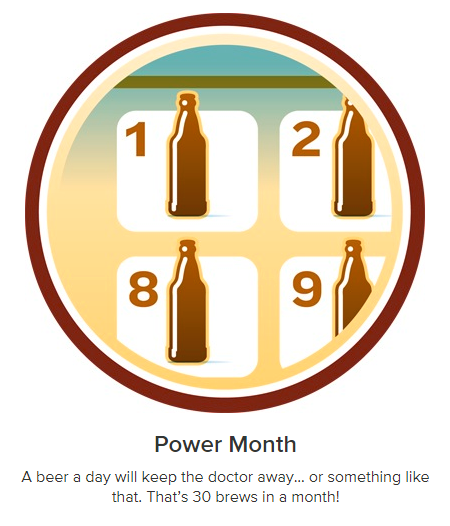}
  \end{minipage}\hfill
  \begin{minipage}[t]{0.3\textwidth}
    \vspace{0pt}
    \includegraphics[width=\linewidth]{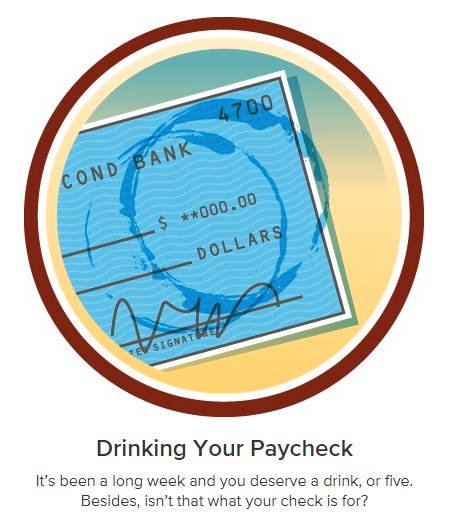}
  \end{minipage}
  \caption{Comparison of three Untappd badges: (a) ABV-based, (b) and (c) frequency consumption-based.}
  \label{fig:badge_comparison1}
\end{figure*}

\subsection{Alcohol-by-Volume (ABV) Badges}

Badges grouped into this category are related to the intensity of alcohol consumption. To obtain them, users must check in a certain number of beers with high alcohol percentage. For example, \textit{Sky's the Limit} badge (Figure \ref{fig:badge_comparison1}a) rewards drinking beers above 10\% ABV, while \textit{Middle of the Road} beers between 5\% and 10\% ABV.

Evaluated through traditional ethical theories. Virtue ethics questions the undermining of self-control when users are rewarded for excess. Utilitarian reasoning suggests the collective harm from normalized high-ABV consumption outweighs any short-term satisfaction. Deontology highlights the problematic manipulation by extrinsic rewards that exploit users' psychological vulnerability.

From a design perspective, ABV badges exemplify a conceptual misalignment: an achievement becomes tied to a risky health behavior. The interface reinforces this by using levels and celebrations of risky behavior. From an EASE perspective, ABV badges compromise autonomy and non-maleficence, illustrating how subtle \textit{dark patterns} normalize harmful behavior when couched in playful design. Although a note now encourages spacing check-ins over time (\textit{`You don't have to do this all in one day, space it out!'}), the core idea remains that stronger equals better.

\subsection{Frequency of Consumption Badges}

Badges in this category, such as \textit{Drinking Your Paycheck} (five drinks in a Friday, see Figure~\ref{fig:badge_comparison1}b) and \textit{Daily Checker} (seven check-ins in consecutive days), effectively gamifying a `streak' of repeated alcohol consumption. This logic is further extended in badges such as \textit{Power Month} (see Figure~\ref{fig:badge_comparison1}b) and \textit{The Usual} (30 and 15 check-ins in a month, respectively), which reward consistent consumption. Even when granted as one-time rewards, their incentive structure turns quantity into accomplishment, subtly encouraging repeated attempts after failure. No observable changes or disclaimers have been made to these badges since 2020.

\begin{figure*}[!ht]
  \centering
  \begin{minipage}[t]{0.3\textwidth}
    \vspace{0pt}
    \includegraphics[width=\linewidth]{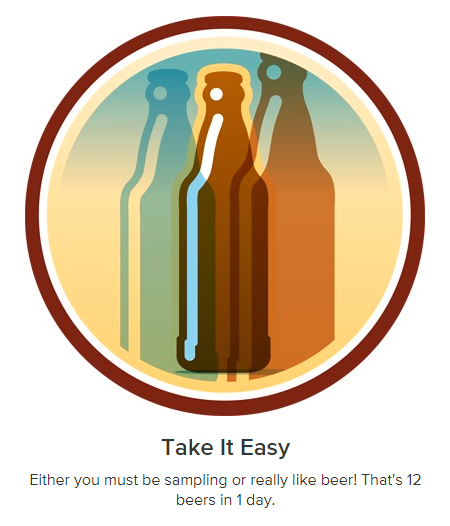}
  \end{minipage}\hfill
  \begin{minipage}[t]{0.3\textwidth}
    \vspace{0pt}
    \includegraphics[width=\linewidth]{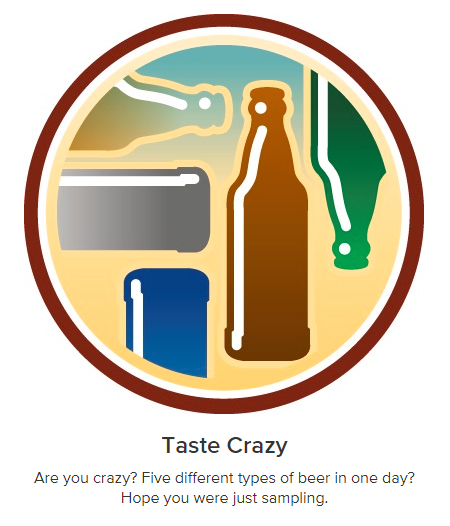}
  \end{minipage}\hfill
  \begin{minipage}[t]{0.3\textwidth}
    \vspace{0pt}
    \includegraphics[width=\linewidth]{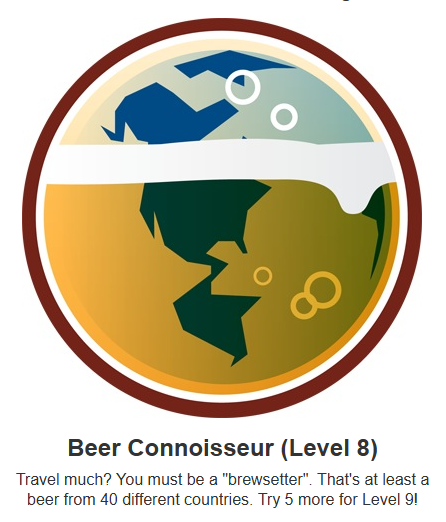}
  \end{minipage}
  \caption{Comparison of three quantity-based badges: (a) and (b) are earned through consumption within a short period, while (c) illustrates an example of a leveled-up badge.}
  \label{fig:badge_comparison2}
\end{figure*}

Streak mechanics frame frequent drinking as a marker of progress, rewarding concentrated or sustained consumption rather than moderation. They encourage habit formation and peer pressure behaviors contrary to responsible consumption norms. From a utilitarian view, the trade-off favors engagement metrics over public health; from a deontological view, the design manipulates users into repeating a behavior for reward. Virtue ethics would question how such mechanisms affect self-discipline and temperance.

Frequency-based badges exemplify a design that prioritizes engagement over user well-being. A conceptual misalignment is noted on the gamification of a `streak' (repeated alcohol consumption) that users would not expect to be celebrated, also noted in community discourse~\cite{noauthor_tonys_2022}. \textit{Dark patterns} emerge from reward loops that conflict with public health or regulatory norms. Such gamification features persist, reflecting a lack of ethical verification.

\subsection{Quantity Badges}

In contrast to the previous category, these badges gamify volume consumption. For example, \textit{Take It Easy} (see Figure~\ref{fig:badge_comparison2}a) ironically rewards checking in 12 beers in one day, while \textit{Taste Crazy} rewards five repeated check-ins of the same beer. These badges remain unchanged from our preliminary analysis.

Other quantity-related badges reward style-specific check-ins (e.g. for IPAs, lagers, stouts, etc.;  see Figure~\ref{fig:badge_comparison2}c). Although less harmful, this scaling mechanism creates a seemingly endless reward loop that nudges users to consume more. In recent observations, these badges are capped at level 100.

Quantity-based badges gamify excess by turning volume consumption into a measurable goal. Interestingly, an independent third-party analytics tool~\cite{hoffman_webbreacheruntappdscraper_2025} even flags potential `binge drinking' events based on the National Institute on Alcohol Abuse and Alcoholism's (NIAAA) definition. The existence of such tools highlights public concern about Untappd's gamification.

From a virtue ethics perspective, repeated engagement can erode temperance and foster excess. A utilitarian view holds that the harms of over consumption outweigh the enjoyment of digital rewards. A deontological lens further raise concerns about manipulation and diminished user autonomy as the system's rely on extrinsic rewards to drive behavior.

Design-wise, this exemplify a conceptual drift~\cite{caragay2024beyond}: the original purpose of achievement (skill mastery) is remapped onto raw consumption. \textit{Dark patterns} emerge by exploiting goal-oriented and completion tendencies. The inclusions of playful disclaimers such as \textit{`Hope you’re just sampling!'} uses irony that blurs ethical intent, functioning more as provocation than discouragement. The persistence of such features indicates a lack of ethical validation, leaving potential societal harms unaddressed.

\pagebreak
\subsection{Period or time of the day Badges}

\begin{figure}[!hb]
  \centering
  \vspace{0pt}
  \includegraphics[width=.5\linewidth]{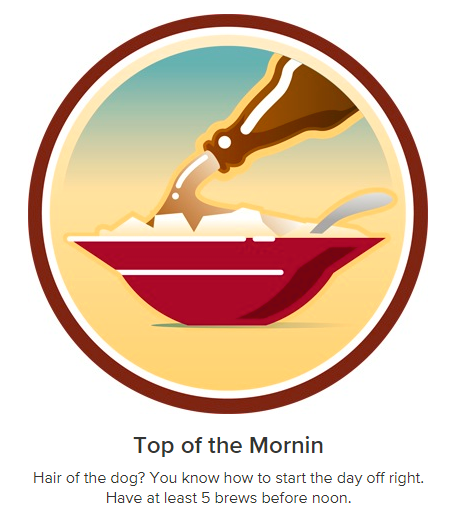}
  \includegraphics[width=.5\linewidth]{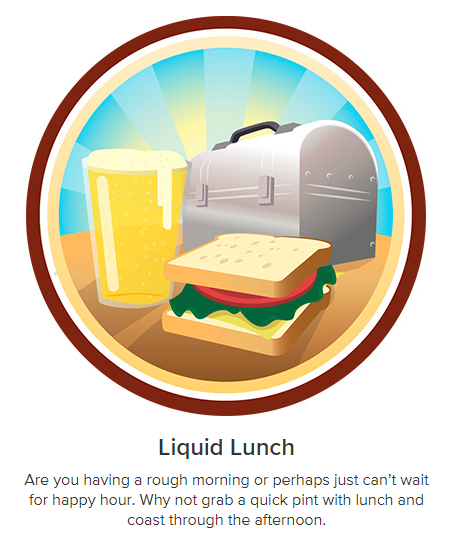}
  \caption{Examples of period-based badges.}
  \label{fig:badge_timeoftheday}
\end{figure}

Badges like \textit{Top of the Mornin} and \textit{Liquid Lunch} (see Figures~\ref{fig:badge_timeoftheday}) are awarded for check-ins at specific hours, encouraging morning or midday drinking. They are earned only once, but users may attempt them repeatedly.

Temporal cues foster habit formation by reinforcing the association between alcohol consumption and daily routines. From a virtue ethics perspective, this pattern conflicts with moderation and self-restraint, effectively promoting vice. Utilitarian reasoning highlights broader public-health risks of normalizing early or habitual drinking, particularly when badges are publicly shared within one’s social network. A deontological analysis further points out the manipulation of user autonomy.

A conceptual misalignment emerges when drinking in the morning or during work hours is framed as socially acceptable or even desirable. In the terms of \textit{dark patterns}, these features reshape socio-normative boundaries through playful metaphors that downplay potential harm. From an EASE perspective, the design shows insufficient ethical validation: stakeholder values such as responsible consumption are neither captured nor reinforced through the feature’s implementation. Simple design changes, like context-sensitive disclaimers or removal of certain triggers, could mitigate these risks; yet such safeguards remain absent.

\subsection{Venue Badges}

\begin{figure}[!h]
  \centering
  \includegraphics[width=\linewidth]{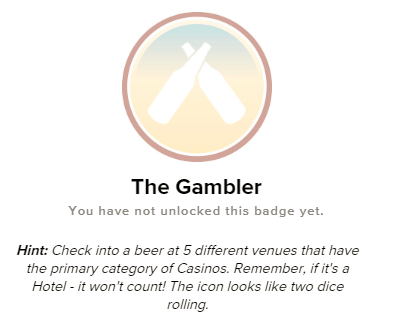}
  \caption{\textit{The Gambler} is a location-based badge; it has not yet been earned by the user so the application hints on how it can be unlocked.}
  \label{fig:badge_gambler}
\end{figure}

These badges tie achievements to specific locations such as pubs, airports, or casinos, as well as outdoor areas like parks, hills, and mountains. Figure~\ref{fig:badge_gambler}c illustrates The Gambler, where the helper text specifies that hotels do not count as casinos, thereby guiding users toward a particular venue type.

Venue badges were initially designed to connect Untappd with local businesses but have also amplified safety and privacy concerns. They may promote drinking in unsafe contexts, such as during travel or outdoor activities. Furthermore, users may share check-ins with time and location data without realizing how easily this information can be tracked~\cite{hoffman_webbreacheruntappdscraper_2025}. Untappd allow users to set their profiles to private~\cite{noauthor_privacy_2023}, hiding venue badges but with limited privact protection beyond that. In contrast, other location-based platforms (e.g., Foursquare, Google Maps) offer more explicit controls for managing check-in history, advertising preferences, and data retention~\cite{patil2012reasons}.

Under virtue ethics, the feature promotes behaviors that conflict with self-control. From an utilitarian perspective, the potential risk of privacy invasion, unsafe drinking, and location-based marketing, outweigh the marginal benefits of user engagement. Deontological ethics raises concerns about autonomy violations, as users' spatial data are collected and exploited without explicit, informed consent, violating individuals' right to self-determination.

These badges illustrate a conceptual drift, in which exploration is reframed as alcohol consumption and presented as an achievement. Travel and drinking merge into a subtle \textit{dark pattern} that fuses social desirability with risky behavior. From an EASE perspective, the geolocation feature shows a lack of ethical articulation and specification, with continuous tracking prioritizing commercial gain over user privacy.

\subsection{Cross-Badge Observations}

Across all badge categories, several recurring patterns reveal the ethical gray areas embedded in Untappd’s gamification design:

\begin{itemize}
    \item Disclaimers such as `drink responsibly' exist but they function more as symbolic than deterrents, as reward mechanisms continue to incentivize risky consumption~\cite{marty_untappd_2020}.
    \item Untappd's growing integration with local businesses and events convert user activity into promotional content or market research~\cite{noauthor_tonys_2022}.
    \item `Badge-chasing' culture~\cite{noauthor_tonys_2022,marty_untappd_2020} foster a sense of belonging around alcohol consumption. Humorous framing further trivializes risk.
    \item The app operates in a \textit{grey area} by embedding promotion within user-generated content, thus bypassing local advertising regulations~\cite{hoffman_webbreacheruntappdscraper_2025}.
\end{itemize}

Combined, these patterns show the need for ethically aware software engineering practices that integrate anticipatory review, privacy-by-design, and public health considerations throughout the software lifecycle.

\section{Discussion}

\subsection*{RQ1. Evolution of Gamification Features}

Our longitudinal comparison reveals that Untappd has evolved technically but not ethically. While new badges and disclaimers have appeared, the platform’s core gaming mechanics (i.e. badges, streaks, and location-based challenges) remain unchanged. The more explicitly problematic badges such as \textit{Drinking Your Paycheck} and \textit{Take It Easy} persist. The inclusion of symbolic disclaimers does not equate to ethical redesign.

The platform continues to grow commercially through local partnerships and sponsored badges, deepening its integration with breweries and events while retaining gamification features that may incentivize harmful behavior. Thus, despite superficial adjustments, the core ethical challenges identified in 2020 remain visible in 2025, highlighting a need for sustained ethical governance throughout the product's lifecycle.

We observed no significant improvement in how the system addresses behavioral risks, user well-being, or regulatory sensitivity. These shortcomings do not appear to shaped not by deliberate malice but the absence of systematic ethical evaluation. The lack of structural change suggests absence of continuous verification or post-deployment validation within the software lifecycle. In other words, while the platform has evolved functionally and expanded commercially, there is little evidence of responsible SE practice.

\subsection*{RQ2. Ethical Interpretations of Gamification in Untappd}

Using our four-dimensional lens (see Section \ref{sec:ethical-framework}), Untappd's  design reveal a pattern of prioritizing engagement over well-being:

\textbf{Individual well-being.} Streaks, ABV badges, and quantity-based achievements encourage excess, habit formation, and dependence. We identified several \textit{conceptual misalignments}, mostly based on the misguided interpretation of alcohol consumption as a mastery. From an EASE perspective~\cite{alidoosti2021ethics}, this suggests a lack of assessment of user well-being impacts during software maintenance and evolution cycles.

\textbf{Social norms.} Badges serve as social signals, and public `toasts' turn individual drinking acts into social validation. This normalizes consumption, raising concerns about promoting a regulated and health-risking substance as cultural practice. Gamification and humor further trivializes risk, disguising potential harm through playfulness.

\textbf{Commercial and regulatory context.} Sponsored badges and venue partnerships blur the line between user engagement and marketing. In many countries, promoting drinking by time or location is regulated, yet social platforms bypass these rules by embedding promotion within user activity. The absence of post-deployment validation allows these practices to persist without recalibration.

\textbf{Philosophical ethics.} From a virtue-ethical view, Untappd foster overindulgent habits. A utilitarian view notes the imbalance between user enjoyment and potential harms to health and safety. Deontologically, gamified features exploit cognitive biases and reduce autonomy. Together, these perspectives show that Untappd's design embodies what the \textit{dark pattern} literature identifies as persuasive yet ethically opaque engagement strategies.

Overall, Untappd operates in an ethically gray zone: not designed with harmful intent, but structured in ways that normalize and reinforce risky behavior. The absence of continuous ethical verification allows these issues to persist.

\subsection{Implications to Practice}

Our analysis of the Untappd case highlights a need to move beyond functionality toward more responsible value-aware design. Several lifecycle-oriented lessons emerged from applying the EASE framework~\cite{aydemir2018roadmap}:

\textbf{Ethical software design.} In socially sensitive domains such as this, engagement metrics alone are insufficient indicators of success. Values such as well-being and autonomy must be articulated early in the design process. Software practitioners should assess how design decisions affect user behavior and long-term well-being.

\textbf{Requirements elicitation and specification.} Ethical requirements should constrain persuasive mechanics and promote moderation rather than excess. Stakeholder values and expectations should be elicited through reflective methods and documented as part of requirements specification. 

\textbf{Building and testing software based on ethics specification.} The articulated values must guide implementation decisions and acceptance criteria. A key challenge is to preventing the de-prioritization of ethical requirements in favor of user retention and time-to-market goals.

\textbf{Privacy and transparency.} Location-based features require ongoing verification of privacy compliance and explicit consent mechanisms. Validation should include transparent feedback loops that inform users how data are collected, shared, and retained.

\textbf{Ethics as an iterative process.} Ethics-aware development demands sustained reflection across the lifecycle. Embedding awareness, conscious valuing, and transparency as continuous enablers ensures that software project goals remains aligned with responsibility and public well-being.

\subsection{Limitations and Threats to Validity}

As with all qualitative and interpretive research, our analysis is subject to several validity threats that must be acknowledged~\cite{given2008sage}.

\textbf{Credibility.} Our findings derived from researcher-driven observations on visible app features and public discourse rather than first-hand user inquiry. This introduces the possibility of selective or interpretive bias. We mitigated these risks through triangulation and reflexive discussion. Also, we built our interpretation on a layered approach combining established ethical principles with practical frameworks for ethical-oriented design.

\textbf{Transferability.} The particularities of Untappd and the cultural context limit the generalization of our findings. Alcohol is a regulated and culturally sensitive substance, and the ethical interpretation may differ across societies. Our results offer analytical interpretations rather than universal claims.

\textbf{Dependability.} The longitudinal nature of our analysis introduces challenges in keeping consistency over time. The scope and granularity of data varied between observations from 2020 and 2025. The platform and its social context also evolved, making direct comparison difficult.

\textbf{Confirmability.} Our interpretations rely on external observation, without access to design artifacts or internal processes. Consequently, our ethical framing may reflect the user perspective rather than those of the platform's designers.

\textbf{Reflexivity.} Ethical reflexivity is particularly important in this domain. Our engagement with Untappd was observational rather than participatory. We acknowledge that the absence of direct community interaction limits the ethnographic depth of our study.

Future studies could extend this work by addressing its limitations: (1) developer-focused triangulation may strengthen credibility, (2) cross-cultural comparisons may also enhance transferability, and (3) access to design documentation would increase confirmability.

\section{Conclusion}

Revisiting our preliminay ethical analysis of Untappd reveals that its gamification remains ethically questionable. Between 2020 and 2025, little structural change occurred beyond new badges and minor interface updates. Core gaming mechanics such as badges, streaks, and location-based challenges persist, continuing to encourage behaviors that raise ethical concerns.  

Untappd's gamification operates in an ethically gray space where community engagement meets normalized risky consumption. The absence of continuous ethical verification within the software lifecycle expose a persistent gap in ethically aware SE practice. 

Gamification should not rely on engagement as its main goal. In experience-driven systems, ethical reflection must be integrated throughout the software process to anticipate social impact, safeguard user autonomy, and align playful design with responsibility.


\bibliographystyle{ACM-Reference-Format}
\bibliography{references}

\end{document}